\documentclass[9pt]{article}
\usepackage{spconf,amsmath,graphicx,hyperref}

\usepackage{cite}
\usepackage{amssymb,amsfonts}
\usepackage{algorithmic}
\usepackage{textcomp}

\usepackage{siunitx} 

\usepackage{booktabs}
\usepackage{color}
\usepackage{url}
\usepackage{multirow}
\usepackage{diagbox}
\usepackage{float}
\usepackage{subfig}
\usepackage{enumitem}
\usepackage{xurl}

\usepackage[table]{xcolor} 
\definecolor{mycolor}{RGB}{230, 230, 255} 

\usepackage{etoolbox}

\apptocmd{\thebibliography}{\setlength{\itemsep}{1pt}\setlength{\parskip}{1pt}\linespread{1}\selectfont}{}{}

\title{Towards Evaluating Generative Audio: \\Insights from Neural Audio Codec Embedding Distances}
\twoauthors
 {Arijit Biswas}
	{Dolby Germany GmbH, Nürnberg, Germany}
 {Lars Villemoes}
	{Dolby Sweden AB, Stockholm, Sweden}
\begin{document}
%
\maketitle
\begin{abstract}

Neural audio codecs (NACs) achieve low-bitrate compression by learning compact audio representations, which can also serve as features for perceptual quality evaluation. We introduce DACe, an enhanced, higher-fidelity version of the Descript Audio Codec (DAC), trained on diverse real and synthetic tonal data with balanced sampling. We systematically compare Fréchet Audio Distance (FAD) and Maximum Mean Discrepancy (MMD) on MUSHRA tests across speech, music, and mixed content. FAD consistently outperforms MMD, and embeddings from higher-fidelity NACs (such as DACe) show stronger correlations with human judgments. While CLAP LAION Music (CLAP-M) and OpenL3 Mel128 (OpenL3-128M) embeddings achieve higher correlations, NAC embeddings provide a practical zero-shot approach to audio quality assessment, requiring only unencoded audio for training. These results demonstrate the dual utility of NACs for compression and perceptually informed audio evaluation.

\end{abstract}

\section{Introduction}
\label{section:Introduction}

As generative audio expands into applications such as content creation and virtual environments, new models have emerged for generating audio from text, visual inputs, or other signals. This growth underscores the need for reliable generative audio quality evaluation methods. The Fréchet Audio Distance (FAD)\cite{FAD} and Maximum Mean Discrepancy (MMD)\cite{CMMD} are two widely used metrics that compare a test signal’s embedding distribution with that of a large reference dataset.

Prior work highlights the importance of embedding choice and reference dataset~\cite{FADTK}. For example, joint audio-text embeddings, such as CLAP~\cite{CLAP} and CLAP LAION~\cite{LAION-CLAP}, perform well for generative music evaluation~\cite{FADTK}. Among neural audio codecs (NACs), EnCodec (EnC)~\cite{EnCodec} and the Descript Audio Codec (DAC)~\cite{DAC} also yield strong acoustic quality prediction~\cite{FADTK}, and NAC embeddings trained on diverse data and higher compression rates have been shown to better align with human judgments of timbre similarity~\cite{TianSony}. However, prediction accuracy varies with the task. PANNs-WaveGram-LogMel embeddings often outperform music-trained alternatives for environmental sound perception~\cite{FAD_environ}. FAD with CLAP LAION Music (CLAP-M) embeddings struggles to evaluate vocals in source separation~\cite{JaffeSourceSep}, and intrusive FAD variants have been proposed for singing voice evaluation~\cite{BereuterSingingVoice}. MMD has also been reported to align well with human perception, even suggesting that FAD may be unnecessary~\cite{KAD}. Overall, results remain inconclusive regarding which distance measure is superior.

This paper presents a systematic study of audio quality prediction using FAD and MMD across three encoder domains: EnC, DAC, and a novel enhanced DAC (DACe) trained with synthetic tonal data. While FAD and MMD are typically used for zero-shot, reference-free evaluation, their efficacy fundamentally depends on whether the underlying embeddings capture human perceptual judgments. To validate this foundational requirement, we first evaluate these metrics on the simpler task of predicting full-bandwidth audio quality given a reference signal and comparing correlations with MUSHRA~\cite{MUSHRA} scores. This controlled, signal-to-signal comparison serves as a necessary validation step: if an embedding space does not reflect human perception in this setting, it is unlikely to succeed in a true reference-free scenario. Furthermore, since high-quality subjective datasets for generative audio are scarce (e.g., MusicEval~\cite{MusicEval}, DCASE 2023 Task 7~\cite{FoleySoundSynth}, and SongEval~\cite{SongEval} are either not full-bandwidth or contain MP3 artifacts), and the choice of reference set is crucial~\cite{FADTK}, our approach of direct signal-to-signal comparisons against a reliable MUSHRA test reduces uncertainty and enables clearer conclusions.

This paper makes the following contributions:
\begin{itemize}[topsep=0pt, parsep=0pt, itemsep=0pt]
\item An enhanced DAC (DACe), trained on a diverse dataset with synthetic tonal audio and novel balanced sampling by coding difficulty to improve performance on challenging tonal material.
\item The first systematic, zero-shot evaluation of NAC embeddings using FAD and MMD, which provides empirical evidence that higher-fidelity codecs yield embeddings more strongly correlated with human ratings. This validates the dual utility of NACs as both compression tools and effective feature extractors for audio quality assessment.
\end{itemize}

The paper is organized as follows: Section~\ref{subsection:trainingset} introduces FAD and MMD; Section~\ref{section:neuralcodecs} describes the NACs under consideration; Section~\ref{section:experiments} presents experimental results; and Section~\ref{section:conclusion} concludes the paper.

\section{Distance Measures}
\label{subsection:trainingset}

We provide a high-level overview of common approaches for evaluating generative audio. The basic idea is to measure the statistical distance between embeddings from the evaluation set and those from a high-quality ``studio reference'' set. Let $\mathbf{X} = {\{x_i\}}_{i=1}^{n}$ and $\mathbf{Y} = {\{y_j\}}_{j=1}^{m}$ denote sequences of embedding vectors from the reference set (distribution \(P\)) and test set (distribution \(Q\)), respectively. Typically, there are multiple embeddings per audio signal, so we generally have $n,m > 1$ even for single signals.  

The FAD~\cite{FAD} is computed from the means $\mu_{\mathbf{X}}, \mu_{\mathbf{Y}}$ and covariances $\Sigma_{\mathbf{X}}, \Sigma_{\mathbf{Y}}$ of the embeddings:  
\[
\mathbf{FAD}^2(\mathbf{X}, \mathbf{Y}) = \|\mu_{\mathbf{X}} - \mu_{\mathbf{Y}}\|_2^2 + \text{tr}\left(\Sigma_{\mathbf{X}} + \Sigma_{\mathbf{Y}} - 2 \sqrt{\Sigma_{\mathbf{X}} \Sigma_{\mathbf{Y}}}\right).
\tag{1}
\]  

This definition assumes a multivariate Gaussian distribution, is sensitive to dataset sample size, and incurs high computational costs. Although $\mathbf{FAD}_\infty$~\cite{FADTK} mitigates sample size bias by extrapolating to infinite samples, other limitations remain. To address these issues, a distribution-free, unbiased, and computationally efficient metric based on MMD~\cite{CMMD} is employed.

Formally, given a kernel function $k(\cdot,\cdot)$, the MMD is defined as  
\begin{equation}
\textstyle
\begin{aligned}
\mathbf{MMD}^2(P,Q) &= \mathbb{E}_{x,x'}[k(x,x')] + \mathbb{E}_{y,y'}[k(y,y')] \\
&\quad - 2\,\mathbb{E}_{x,y}[k(x,y)],
\end{aligned}
\tag{2}
\end{equation}   
where \(x, x'\) are drawn from \(P\) and \(y, y'\) from \(Q\).

For finite samples from the reference set \(\mathbf{X} = \{x_i\}_{i=1}^{n}\) and the test set \(\mathbf{Y} = \{y_j\}_{j=1}^{m}\), an unbiased estimator of Eq.~(2) is given by  
\begin{equation}
\textstyle
\begin{aligned}
\widehat{\mathbf{MMD}}^2 = & \frac{1}{n(n-1)}\sum_{i\neq j} k(x_i,x_j) 
+ \frac{1}{m(m-1)}\sum_{i\neq j} k(y_i,y_j) \\
& - \frac{2}{nm}\sum_{i,j} k(x_i,y_j).
\end{aligned} \tag{3}
\end{equation}  
Finally, we define the scaled MMD as:  
\[
\mathbf{MMD}_{s} = \alpha \cdot \widehat{\mathbf{MMD}}^2,
\tag{4}
\]  
where \(\alpha\) is a scaling factor (set by default to 1000). The distance is computed using the Gaussian radial basis function (RBF) kernel~\cite{CMMD, KAD},  
\( k(\mathbf{x}, \mathbf{y}) = \exp\Big(-\frac{\|\mathbf{x} - \mathbf{y}\|^2}{2\sigma^2} \Big), \)
with \(\sigma\) as the bandwidth parameter, determined via a median-distance heuristic~\cite{gretton2012kernel} to effectively discriminate between embedding distributions.  

Although both distances are typically used to evaluate generative audio by comparing embedding distributions of a test set and a reference set, they can also be applied to compare NAC embeddings of a single test signal with a reference signal. For a single signal pair, the number of embedding vectors, $n$ and $m$, remains greater than 1, as multiple embeddings are extracted per signal, ensuring a valid evaluation of Eq.~(4).

\section{Neural Audio Codecs (NACs)}
\label{section:neuralcodecs}

The goal of an audio codec is to compress audio to a low bitrate while reconstructing a signal that is perceptually close to the original. Time-domain, end-to-end neural audio codecs (NACs)~\cite{SoundStream, EnCodec, DAC} typically employ a fully convolutional encoder that processes the waveform and outputs embeddings at a reduced sampling rate. These embeddings are quantized using a residual vector quantizer (RVQ)~\cite{SoundStream}, and a fully convolutional decoder reconstructs an approximation of the original waveform from the quantized embeddings. The model is trained on a diverse audio dataset using a combination of reconstruction and adversarial losses. One or more discriminators are jointly trained to distinguish the original audio from the reconstructed version and to provide a feature-based reconstruction loss. The RVQs are trained end-to-end along with the rest of the model. For full-bandwidth general audio, the following NAC encoders were considered:

\noindent \textbf{EnCodec (EnC):} The 48 kHz EnC NAC~\cite{EnCodec_Github} is a stereo model trained exclusively on music. 

\begin{figure*}[t]
    \centering
    \begin{minipage}[t]{0.48\textwidth}
        \vspace{0pt} 
        \centering
        \includegraphics[width=0.85\linewidth]{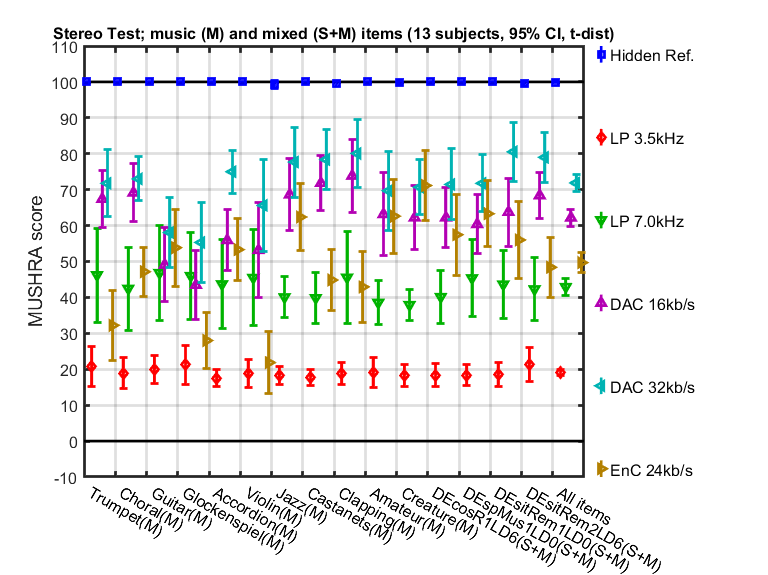}
        \caption{DAC (16 kb/s) outperforms EnC (24 kb/s).}
        \label{fig:dacVSenc}
    \end{minipage}
    \hfill
    \begin{minipage}[t]{0.48\textwidth}
        \vspace{0pt} 
        \centering
        \includegraphics[width=1.0\linewidth]{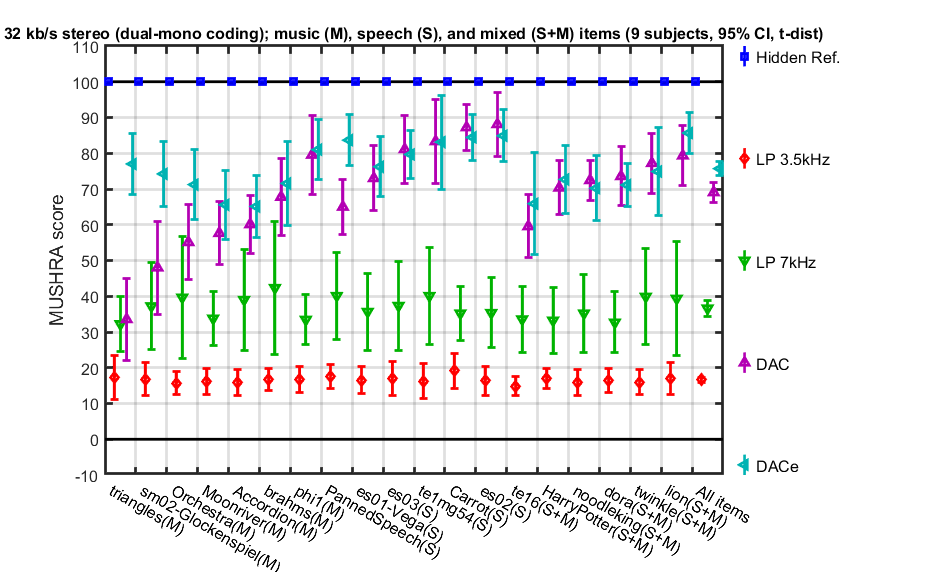}
        \caption{DACe outperforms DAC.}
        \label{fig:dacVSdace}
    \end{minipage}
    
    \vspace{-0.4cm} 
\end{figure*}

\noindent \textbf{Descript Audio Codec (DAC):} The DAC targets general audio coding at 44.1 kHz. We evaluated the publicly available 16 kb/s model~\cite{Descript_16kbps} using a MUSHRA listening test (Fig.~\ref{fig:dacVSenc}), which confirmed that DAC outperforms EnC. A stereo listening test was conducted over headphones with 15 stereo excerpts (48 kHz) from the ODAQ set~\cite{odaq}. Music excerpts were retained, while six dialogue-heavy excerpts were excluded to ensure a fair comparison with EnC. DAC was tested at 16 and 32 kb/s, encoding the left and right channels independently (i.e., dual-mono coding). Input signals were resampled to 44.1 kHz for coding and back to 48 kHz for listening tests. Results show that DAC at 16 kb/s outperformed EnC at 24 kb/s. However, DAC struggles with tonal content, such as glockenspiel, a limitation also reported by its authors~\cite{DAC}.

\noindent \textbf{Enhanced Descript Audio Codec (DACe):} We developed DACe by improving DAC and training it with up to 32 10-bit codebooks, corresponding to $\sim$30~kb/s per channel at 48~kHz. DACe was trained using the same hyperparameters as DAC, but with a batch size of 48 instead of 72, on a diverse 720-hour, 48~kHz mono dataset~\cite{MDCTNet} covering 18 music genres, speech, and isolated instruments. To address DAC’s weakness on tonal material, we added 32 hours of real tonal audio and, more critically, generated synthetic tonal signals on the fly during training to expose the time-domain codec to challenging content. Each synthetic tonal excerpt was created by dynamically simulating multiple tonal events per sample, with frequency, amplitude, and decay evolving according to randomized onset timing and Poisson-distributed event density. Balanced sampling ensured that 33\% of each mini-batch consisted exclusively of challenging synthetic tonal excerpts. A MUSHRA test (Fig.~\ref{fig:dacVSdace}) compared DAC and DACe at 32~kb/s stereo using dual-mono coding. Test excerpts were drawn from an internal dataset (since ODAQ excerpts were not sufficiently challenging), with none overlapping the training set, and were selected based on maximal Mel-loss~\cite{DAC} deviation between the two NACs. As shown in Fig.~\ref{fig:dacVSdace}, DACe consistently outperforms DAC, with strong gains on tonal items (e.g., triangles and the MPEG glockenspiel test file sm02) and on a panned speech excerpt. Overall, DACe outperforms DAC, and both surpass EnC.

For all three NACs, embeddings were extracted from the final encoder layer, prior to RVQ. In the following, we examine how distance measures computed in these embedding domains correlate with subjective evaluations.

\section{Experimental Results}
\label{section:experiments}

\subsection{FAD and MMD in NAC embedding domain}
\label{subsection:FAD-MMD_codec}

For computing the FAD distances, we used the FADTK toolkit~\cite{FADTK_github}. For MMD, we ported the original code~\cite{MMD_github} into this toolkit and implemented a median heuristic~\cite{gretton2012kernel} for kernel bandwidth selection (bandwidth set to the median of pairwise Euclidean distances).

We used MUSHRA tests for benchmarking. Specifically, we correlated these distance measures with subjective scores from the mono MPEG Unified Speech and Audio Coding (USAC) verification listening tests~\cite{usac_lt, USAC}, which included 24 excerpts (8 each of speech, music, and mixed content) coded with USAC, HE-AAC, and AMR-WB+ across bitrates ranging from 8 kb/s to 24 kb/s. The test involved 66 listeners from various organizations. We included only the mono test in our study, as most publicly available NACs and popular embedding models do not fully account for stereo perception. In addition to being one of the largest listening tests conducted in MPEG, the USAC mono test was selected (instead of the ODAQ~\cite{odaq} stereo test) because it provides a balanced mix of speech, music, and mixed excerpts, and the tested codecs were optimally tuned. Prediction accuracy was evaluated using the Pearson correlation ($R_p$) to measure linearity and the Spearman correlation ($R_s$) to measure monotonicity, with higher values indicating better performance. As shown in Table~I, where the best-performing models are highlighted in bold, distances computed in the EnC embedding domain exhibit the weakest correlation with subjective data. Correlation improves significantly when distances are computed in the DAC embedding domain, supporting the hypothesis that higher-fidelity NACs learn more robust features for quality prediction tasks. The table also includes the DAC 8 kb/s model~\cite{Descript_8kbps}, as this variant is a common choice in generative audio applications.

Further improvement is observed with DACe, indicating stronger alignment with subjective judgments. This likely stems from DACe's optimization, which encourages its latent representation to retain perceptually salient information while discarding inaudible details. This produces a compact embedding space in which distances reliably reflect perceptual quality degradations and remain robust to irrelevant variations.

Between the two distance measures, we observe that FAD correlates more strongly with subjective perceptual scores. From Eqs.~(1)--(4), the squared Fréchet Audio Distance, $\mathbf{FAD}^2(\mathbf{X},\mathbf{Y})$, depends solely on the empirical means and covariances of the embeddings. In contrast, the unbiased MMD estimator, $\widehat{\mathbf{MMD}}^2$, is a U-statistic sensitive to higher-order moments via a kernel $k$~\cite{gretton2012kernel}.

We hypothesize that FAD's empirical advantage arises from two factors. First, by relying on the first two moments, its estimator yields lower finite-sample variance and greater stability for embedding spaces derived from time-domain NACs, under an assumption of near-Gaussianity. In contrast, MMD, although distribution-free and computationally efficient~\cite{CMMD,KAD}, has variance that is highly sensitive to kernel choice and bandwidth~\cite{grettonNIPS2012}, which can amplify noise or irrelevant differences. While advanced kernel selection is beyond the scope of this work, experiments with DACe embeddings showed that the median-heuristic bandwidth is around 73. Fixed bandwidths of $\sigma = 1, 10, 100, 1000, 10000$ were also tested. Correlation with subjective data peaked near $\sigma = 100$ but remained lower than with the median-distance heuristic. The drop was sharper when decreasing $\sigma$ toward 1 (kernel overly sensitive) than when increasing $\sigma$ toward 10000 (kernel too insensitive), indicating that the median heuristic provides an effective bandwidth choice. These findings are consistent with prior reports (see~\cite{KAD}, Appendix A). Second, human auditory perception is primarily sensitive to spectral balance, temporal envelope, and overall dynamics~\cite{moore2012introduction}, which can be captured by FAD through low-order statistics. Taken together, these factors make FAD a direct and stable measure of perceived audio quality, requiring no additional hyperparameter tuning beyond the choice of embedding model.

\begin{table}[t]
\setlength\tabcolsep{5pt} 
\centering
\caption{Performance of FAD and MMD across neural codec and popular audio embedding models, showing Pearson ($R_p$) and Spearman ($R_s$) correlations with MUSHRA scores. Embedding dimension (Dim) and sample rate (SR, in kHz) are indicated. ‘w/o LP’ refers to results excluding lowpass anchor conditions. Best results are highlighted in bold.}
\scriptsize
\begin{tabular}{|l|r|S[table-format=3.1]|S[table-format=1.2]|S[table-format=1.2]||S[table-format=1.2]|S[table-format=1.2]|}
\hline
\textbf{Encoder} & \textbf{Dim} & \textbf{SR} & \multicolumn{2}{c||}{\textbf{All conditions}} & \multicolumn{2}{c|}{\textbf{w/o LP}} \\ \cline{4-7}
& & & {$\mathbf{R_p}\uparrow$} & {$\mathbf{R_s}\uparrow$} & {$\mathbf{R_p}\uparrow$} & {$\mathbf{R_s}\uparrow$} \\ 
\hline\hline

\multicolumn{7}{|c|}{\cellcolor{gray!15}\textbf{Neural audio codec embeddings}} \\ \hline
\rowcolor{gray!10}\multicolumn{7}{|l|}{\textit{\textbf{MMD}}} \\ 
EnC             & 128  & 48.0 & 0.41 & 0.70 & 0.31 & 0.65 \\ 
DAC 8 kb/s          & 1024 & 44.1 & 0.62 & 0.76 & 0.54 & 0.69 \\
DAC 16 kb/s         & 128  & 44.1 & \textbf{0.65} & \textbf{0.77} & 0.57 & 0.69 \\ 
\rowcolor{blue!10} DACe 24 kb/s        & 1024 & 48.0 & \textbf{0.65} & \textbf{0.77} & \textbf{0.60} & \textbf{0.71} \\ 
\rowcolor{gray!10}\multicolumn{7}{|l|}{\textit{\textbf{FAD}}} \\ 
EnC             & 128  & 48.0 & 0.38 & 0.66 & 0.32 & 0.63 \\ 
DAC 8 kb/s          & 1024 & 44.1 & 0.67 & 0.80 & 0.61 & 0.74 \\ 
DAC 16 kb/s         & 128  & 44.1 & 0.68 & 0.81 & 0.65 & 0.75 \\ 
\rowcolor{blue!20} DACe 24 kb/s        & 1024 & 48.0 & \textbf{0.70} & \textbf{0.82} & \textbf{0.69} & \textbf{0.77} \\ 
\hline\hline

\multicolumn{7}{|c|}{\cellcolor{gray!15}\textbf{Popular audio embeddings}} \\ \hline
\rowcolor{gray!10}\multicolumn{7}{|l|}{\textit{\textbf{MMD}}} \\ 
CLAP                & 1024 & 44.1 & 0.54 & 0.47 & 0.57 & 0.51 \\ 
CLAP-A    & 512  & 48.0 & 0.73 & 0.77 & 0.70 & \textbf{0.78} \\ 
\rowcolor{blue!10} CLAP-M    & 512  & 48.0 & \textbf{0.76} & \textbf{0.80} & 0.67 & 0.74 \\ 
OpenL3-128E   & 512  & 48.0 & 0.74 & 0.72 & 0.70 & 0.69 \\ 
OpenL3-128M   & 512  & 48.0 & 0.73 & 0.75 & 0.72 & 0.75 \\ 
OpenL3-256E   & 512  & 48.0 & 0.58 & 0.57 & 0.66 & 0.65 \\ 
OpenL3-256M   & 512  & 48.0 & 0.59 & 0.59 & \textbf{0.79} & \textbf{0.78} \\ 
\rowcolor{gray!10}\multicolumn{7}{|l|}{\textit{\textbf{FAD}}} \\ 
CLAP                & 1024 & 44.1 & 0.54 & 0.52 & 0.58 & 0.57 \\ 
CLAP-A    & 512  & 48.0 & 0.76 & 0.79 & 0.74 & 0.80 \\ 
\rowcolor{blue!20} CLAP-M    & 512  & 48.0 & \textbf{0.85} & \textbf{0.88} & 0.82 & 0.85 \\ 
OpenL3-128E   & 512  & 48.0 & 0.83 & 0.83 & 0.83 & 0.82 \\ 
OpenL3-128M   & 512  & 48.0 & 0.84 & 0.84 & \textbf{0.86} & \textbf{0.86} \\ 
OpenL3-256E   & 512  & 48.0 & 0.61 & 0.57 & 0.74 & 0.73 \\ 
OpenL3-256M   & 512  & 48.0 & 0.68 & 0.67 & 0.65 & 0.65 \\ 
\hline
\end{tabular}
\end{table}

\vspace{-1em}

\subsection{FAD and MMD in popular embedding domain}
\label{subsection:FAD-MMD_emd}

In this section, we repeated the above experiment using popular full-bandwidth audio embedding models (44.1 or 48 kHz sample rate): CLAP, CLAP LAION (both audio-trained CLAP-A and music-trained CLAP-M variants), and OpenL3~\cite{OpenL3, OpenL3_github}. Our goal was to identify which model best predicts audio quality. The OpenL3 models tested include variants trained on environmental sounds (OpenL3-128E, OpenL3-256E) and music (OpenL3-128M, OpenL3-256M) at 128- and 256-band Mel-spectrogram resolutions. To this end, we integrated only the OpenL3 embeddings into the FADTK toolkit, while the other embedding models were already included. For OpenL3, we considered only the 512-dimensional embeddings used in~\cite{JordiOpenL3}, excluding the 6144-dimensional version.

From Table~I, FAD with CLAP-M performs best, followed by OpenL3-128M, while the standard CLAP model performs worst.

CLAP-M likely outperforms other embeddings due to its large (estimated $\sim$11,000 hours), diverse training data and language-grounded contrastive objectives, which align perceptually relevant audio cues with text and make FAD distances in this space a strong proxy for perceived quality. OpenL3, trained on audio–visual correspondence, captures sound identity but lacks linguistic grounding, explaining its second-best performance, while the standard CLAP model, trained on smaller datasets, yields less generalizable quality representations.

Overall, CLAP-A/M and OpenL3-128 models outperform the best NAC embeddings (DACe) in both Pearson and Spearman correlations with MUSHRA scores under all conditions. NACs likely underperform due to two key reasons: (i) \textbf{training objective mismatch,} as NACs are optimized for reconstruction and compression, while CLAP and OpenL3 use contrastive or self-supervised objectives that inherently capture perceptual similarity and semantics; and (ii) \textbf{training data scale and diversity,} since popular audio embeddings are trained on massive, heterogeneous datasets, whereas NACs like DACe are trained with roughly ten times less data.

\section{Conclusion}
\label{section:conclusion}

We presented a systematic study of audio quality prediction using FAD and MMD across NAC and popular embedding domains in a full-reference setting. Our results demonstrate that FAD consistently outperforms MMD in correlating with human judgments. Distances computed in pre-trained NAC embeddings align well with subjective scores and improve with codec fidelity. This indicates that high-fidelity NACs have the potential to serve as zero-shot feature extractors for generative audio quality assessment without large-scale labeled datasets. Popular embeddings, such as CLAP LAION Music (CLAP-M) and OpenL3 Mel128 (OpenL3-128M), achieve higher correlations, likely due to larger, more diverse training datasets and contrastive or self-supervised objectives that capture perceptually relevant features. Nevertheless, NAC embeddings uniquely combine compression and perceptual evaluation in a single model and require only unencoded audio for training. These findings motivate future work on developing higher-fidelity NACs with expanded datasets and alternative training objectives or architectures that better align their embeddings with human perception.

\newpage

\bibliographystyle{IEEEbib}
\bibliography{strings,refs}

\end{document}